\documentclass[12pt,preprint]{aastex}

\slugcomment{accepted for publication in AJ}

\shorttitle{SMA observation of NGC~7027}
\shortauthors{Nakashima et al.}

\begin{document}


\title{Three-Dimensional Structure of the Central Region of NGC~7027: A Quest for Trails of High-Velocity Jets}


\author{Jun-ichi Nakashima, Sun Kwok and Yong Zhang}
\affil{Department of Physics, University of Hong Kong, Pokfulam Road, Hong Kong\\ Email(JN): junichi@hku.hk}

\author{Nico Koning}
\affil{Department of Physics and Astronomy, University of Calgary, Calgary, Canada T2N 1N4}


\begin{abstract}
We report on the results of a radio interferometric observation of NGC~7027 in the CO $J=2$--1 and $^{13}$CO $J=2$--1 lines. The results are analyzed with morpho-kinematic models developed from the software tool {\it Shape}. Our goal is to reveal the morpho-kinematic properties of the central region of the nebula, and to explore the nature of unseen high-velocity jets that may have created the characteristic structure of the central region consisting of molecular and ionized components. A simple ellipsoidal shell model explains the intensity distribution around the systemic velocity, but the high velocity features deviate from the ellipsoidal model. Through the {\it Shape} automatic reconstruction model, we found a possible trail of a jet only in one direction, but no other possible holes were created by the passage of a jet.
\end{abstract}




\section{Introduction}

Although many planetary nebulae (PNe) have simple apparent morphologies, their true 3-D structures are in fact not well known.  This is true even for the most well-observed PN NGC 7027.  In the optical and radio images, it has a clear shape of a ring, similar in appearance to the Ring Nebula NGC 6720. Its optical and radio continuum structures, corresponding to the ionized gas component of the nebula, have been analyzed by various authors, generally suggesting that it is an ellipsoidal shell inclined at $\sim30^{\circ}$ to the line of sight \citep{sco73,mas89,roe91,wal94}. Observations in the molecular hydrogen (H$_2$) line, however, reveal that it has a bipolar structure \citep{lat00}. There is also evidence that there may be multiple collimated jets in the nebula \citep{cox97}. 

NGC 7027, being a very young PN \citep[dynamical age~$\sim$~600 years;][]{mas89}, still possesses the remnant of the molecular envelope from the asymptotic giant branch (AGB) progenitor.  Its CO envelope was first detected by \citet {muf75} and has been imaged by single-dish telescopes and interferometric techniques \citep{mas85,bie91,jam91,gra93,fon06}.  In addition to high spatial resolution, the molecular-line interferometric observations have the additional advantage that the high spectral resolution observations provide the additional dimension of kinematic structure. If interpreted by a proper analytical tool, interferometric data can provide powerful insights into the true 3-D structure of the nebula.

In this paper, we report on the result of the millimeter-wave interferometric observation of NGC~7027 in the CO $J=2$--1 and $^{13}$CO $J=2$--1 lines using the Submillimeter Array\footnote{The Submillimeter Array is a joint project between the Smithsonian Astrophysical Observatory and the Academia Sinica Institute of Astronomy and Astrophysics, and is funded by the Smithsonian Institution and the Academia Sinica.} (SMA). These observational results are analyzed with a morphokinematic modeling tool {\it Shape} \citep{ste06}. Our purpose is to reveal the structure of the central region consisting of molecular and ionized components and to find possible trails of jets that might have passed through the central region.

The outline of this paper is as follows. In Section~2 we describe the details of the observation and data reduction. In Section~3 we present the observational results. In Section~4 we present the procedure and results of the morphokinematic modeling with {\it Shape}. In Section~5 we discuss the obtained results, and finally the present research is summarized in Section~6.


\section{Observation and Data Reduction}
The radio interferometric observation of NGC~7027 was made on September 20, 2004 with the SMA. The SMA data set was downloaded from the Radio Telescope Data Center (RTDC) operated by the Harvard Smithsonian Center for Astrophysics. The data were obtained under good atmospheric conditions with a zenith optical depth at 230 GHz of 0.045--0.055. The two CO lines (i.e., CO $J=2$--1 and $^{13}$CO $J=2$--1 at 230.538000~GHz and 220.398681~GHz, respectively) were included in the observed frequency ranges of 229.4--231.3~GHz (USB) and 219.4--221.3~GHz (LSB). The array used $6\times6$~m elements in the compact array configuration. The baseline length ranged from 10 to 70~m. The field of view of a single antenna was about 54$''$. The observation was interleaved every 20 minutes with nearby gain calibrators, 2202+422 and 2015+371, to track the phase variations over time. We calibrated the data using IDL--MIR, which is a data reduction software package developed by the SMA project. The absolute flux calibration was determined from observations of Uranus, and was approximately accurate to within 15\%. We followed a standard calibration method. The final map has an accumulated on-source observing time of about 6 hours. The single-sideband system temperature ranged from 90 to 120~K, depending on the atmospheric conditions and airmass. The SMA correlator had a bandwidth of 2~GHz with a resolution of 0.812~MHz. The velocity resolution was roughly 1.06 km~s$^{-1}$ at the observed frequency. The phase center of the map was taken at R.A.$=21^{\rm h}07^{\rm m}0.1589^{\rm s}$, decl.$=+42^{\circ}14'10.169''$ (J2000). Image processesing of the data was performed with the MIRIAD software package \citep{sau95}. The robust weighting gave a 2.74$''$ $\times$ 1.37$''$ CLEAN beam with a position angle of $75.9^{\circ}$. 

We detected the continuum emission by integrating over a 1.0 GHz range in USB (using emission free channels), and the flux density of the emission is measured to be 1.3 Jy. This value is significantly smaller than previous measurements at (or close to) 230~GHz \citep[3--4 Jy; see, e.g.,][]{haf08}. This is presumably because the extended (dust) component of the continuum emission is resolved out by interferometer. Since the main concern in the paper is the molecular component of the central region, of which the flux is completely recovered, we do not go into specifics about the continuum emission.


\section{Results}
The total flux profiles of the $^{12}$CO $J=2$--1 and $^{13}$CO $J=2$--1 lines are presented in Figure~1. The peak intensity of the $^{12}$CO $J=2$--1 and $^{13}$CO $J=2$--1 lines are 49.1~Jy and 5.3~Jy, respectively. The line widths (at the 0 intensity level) of the $^{12}$CO $J=2$--1 and $^{13}$CO $J=2$--1 lines are 46~km~s$^{-1}$ and 62~km~s$^{-1}$, respectively. The velocity integrated intensities of the $^{12}$CO $J=2$--1 and $^{13}$CO $J=2$--1 lines are 1029.6~Jy~km~s$^{-1}$ and 165.0~Jy~km~s$^{-1}$, respectively. Compared to a single-dish observation with the SMT 10~m telescope \citep{zha08}, 21~\% and 81~\% of the single-dish fluxes in the $^{12}$CO $J=2$--1 and $^{13}$CO $J=2$--1 lines have been recovered respectively in the present interferometric observations. Taking into account the uncertainty in the flux measurements, we could consider that almost all of the single-dish flux of the $^{13}$CO $J=2$--1 line has been recovered, although a significant fraction of the $^{12}$CO flux is lost. The majority of the missing flux of the $^{12}$CO emission must originate in the extended spherical envelope \citep{mas85,fon06}. Given the low optical depth of the $^{13}$CO line, the $^{13}$CO flux must primarily come from the dense, small region surrounding the central ionized region, and we recovered the $^{13}$CO flux almost totally. For this reason, we believe that the $^{12}$CO flux emitted from the molecular component of the central region, which is the main concern of this research, must be almost fully recovered also. Note that in this paper the ''central region'' means the region consisting of the central ionized region and surrounding molecular component that is traced by the $^{13}$CO emission. 

Figure~2 shows the total-flux intensity maps of the $^{12}$CO $J=2$--1 and $^{13}$CO $J=2$--1 lines, superimposed on the $L$-band ($\lambda=18$--20~cm) continuum image \citep{bai03}. It is obvious that the ionized region is embedded inside the molecular emission region. The CO feature appears have an axial symmetric structure with the symmetry axis lying along the northwest to southeast direction (position angle$\sim$135$^{\circ}$). The intensity of the CO emission is weaker in the northwest and southeast edges, and the emission region of the $L$-band continuum spills into this CO-faint regions. This appearance is consistent with previous observations in the CO $J=1$--0 line \citep[see, e.g., Figure~11 in][]{bie91}. The weak features seen in the corners of the upper panel of Figure~2 are uncleanable features, which are due presumably to an extended envelope. The $^{12}$CO $J=2$--1 feature is slightly larger than the $^{13}$CO $J=2$--1 feature in the northwest to southeast direction. The angular size of the emission region of the $^{12}$CO $J=2$--1 line is roughly 26$''$ (major axis: northwest to southeast) and 21$''$ (minor axis: northeast to southwest) at a 3$\sigma$ level. This corresponds to the linear sizes of $3.8\times10^{17}$~cm and $3.1\times10^{17}$~cm at a distance of 980~pc \citep{zij08}. Similarly, the angular size of the emission region of the $^{13}$CO $J=2$--1 line is roughly 22$''$ (major axis) and 18$''$ (minor axis), corresponding to $3.2\times10^{17}$~cm and $2.6\times10^{17}$~cm at a distance of 980~pc.

Figures~3 and 4 show velocity channel maps of the $^{12}$CO $J=2$--1 and $^{13}$CO $J=2$--1 lines, respectively. The emission regions seen in Figures~3 and 4 are very similar to the $^{12}$CO $J=1$--0 maps \citep{mas85,bie91,jam91,gra93,fon06}, even though the spherical component is missing in the present $^{12}$CO $J=2$--1 and $^{13}$CO $J=2$--1 maps. As in previous CO $J=1$--0 maps, the $^{12}$CO $J=2$--1 and $^{13}$CO $J=2$--1 features exhibit a systematic motion: the northwest part of the feature is bright in low velocity channels, whereas the southeast part of the feature is bright in the high-velocity channels. The position--velocity diagram giving in Figure~5 is also consistent with the $^{12}$CO $J=1$--0 observations \citep[see, e.g.,][]{gra93}, which has been interpreted by an expanding ellipsoidal shell.


\section{Morpho-Kinematic Modeling with {\it Shape}}

Since the flux of the central region is almost totally recovered both in the $^{12}$CO $J=2$--1 and $^{13}$CO $J=2$--1 lines, the present interferometric data should be useful to derive the morpho-kinematic properties of the molecular component of the central region. We have constructed models using the {\it Shape} software originally developed by \citet{ste06} for the analysis of optical imaging spectroscopy data of planetary nebulae. In this paper, we have applied the software for radio interferometric data. Although it does not include hydrodynamical or radiation transfer calculations, {\it Shape} is useful to create synthetic images at different velocity channels for comparison with actual observations. {\it Shape} builds a three-dimensional morpho-kinematic model from the distribution of moving particles. Therefore, the {\it Shape} modeling is potentially valid assuming ''optically thin''. Although the $^{12}$CO line is often optically thick, several studies of NGC~7027 show that $^{12}$CO $J=1$--0 line is not very optically thick, i.e., $\tau<1$, and the intensity is approximately proportional to column density \citep{deg90,jam91}. Additionally, we used the data of the $^{13}$CO $J=2$--1, which is usually optically thin, for the analyses in conjunction with the data of the $^{12}$CO $J=1$--0 line. Therefore, the assumption about the optical thickness in the {\it Shape} model does not seem to have harmful effects on the results.

Although {\it Shape} does not directly handle the concept of ''excitation'', what {\it Shape} is actually doing is equivalent to assuming constant excitation (in the scheme of the radiation transfer calculation). Therefore, practically we are assuming, in the {\it Shape} modeling, that the structure of CO gas (of the central region) is fully traced by the present CO observation.

{\it Shape} provides two different ways to construct a model: ``interactive modeling'' and ``automatic reconstruction''. In the interactive modeling, we begin by creating an initial structure and velocity field for the nebula. Synthetic $p$--$v$ diagrams and channel maps are then produced from the model which are subsequently compared with the actual observations. The geometry and velocity structure are then interactively modified until satisfactory convergence is achieved. In the automatic reconstruction, {\it Shape} automatically determines the spatial distribution and velocity of particles by directly scanning the observational maps without any assumption about the nebular structure (note that, however, we need to assume the velocity law). 

In this work, we modeled the central region using both of the aforementioned methods. In the interactive modeling mode, we assume an ellipsoidal shell as the structure of the central region, as in previous work to interpret the $^{12}$CO $J=1$--0 maps \citep[see, e.g.,][]{gra93}. We also use the automatic reconstruction to explore the possible three-dimensional morphology of the nebula. The details of each model are given in the subsections below.

\subsection{Ellipsoidal Shell Model with {\it Shape} Interactive Modeling}

The values of the parameters of the ellipsoidal shell (inclination angles, major- and minor-axis radii, and systematic velocity) are summarized in Table~1. For the kinematic structure, we assume a constant velocity field, as is the case for the kinematic model of the spectra-imaging data of the Br$\gamma$ emission by \citet{cox02}. The expansion velocity we adopt (22~km~s$^{-1}$) is higher than the value (14~km~s$^{-1}$) given by \citet{gra93}. The expansion velocity given by \citet{gra93} was not fast enough in the present case, and therefore we increased it to 22~km~s$^{-1}$, which fits the present observation better. A possible reason for the discrepancy of the expansion velocities between the present work and \citet{gra93} is that the present CO $J=2$--$1$ observation is more sensitive for high velocity regions than previous CO $J=1$--$0$ observations. The expansion velocity, 22~km~s$^{-1}$, is consistent with the line-width of the CO $J=2$--$1$ line. The volume of the ellipsoidal shell is sampled with 20,000 particles, which creates a sufficiently smooth map for comparison with the observation. A circular beam with a diameter of 2$''$ is used to convolve the modeled image to simulate the observations. The velocity channel maps of the model is given in Figure~6.

This simple ellipsoidal model, of course, cannot perfectly reproduce the observations. However, through this simple modeling we can still see that the observed features near the systematic velocity can be well approximated by an expanding ellipsoidal shell.
The deviation of the observed high-velocity components from the ellipsoidal shell model may be related to the possible existence of high-velocity flows (discussed later in Section 5).

\subsection{3D Model with {\it Shape} Automatic Reconstruction}

In addition to the ellipsoidal-shell model, a full 3-D model using the {\it Shape} automatic reconstruction is also generated. The particle distribution of this model is derived from direct scanning of the observed position-velocity diagrams and channel velocity maps. For the velocity field, we assume a linear velocity law $v(r)=2.0r''$ where $v$ is the expanding velocity and $r$ is the distance from the central star. We determined the proportionality constant (i.e., 2.0) so as to produce the most symmetric 3-D structure. We found that a constant velocity (as assumed in the ellipsoidal-shell model) is unable to fit as well as a linear-velocity law. Although one might think that the Hubble-type velocity law suggests linear, continuous acceleration of the envelope motion with a distance from the central star, the velocity law can be also explained by ballistic movements, due to a sudden acceleration in the past plus free expansion since then \citep[see, e.g.,][]{buj08}. Volumes of 50,000 particles each are needed to create smooth maps for comparisons with the $^{12}$CO and $^{13}$CO observations.  A circular beam with a diameter of 2$''$ is used for convolution with the model images.  

The channel-velocity maps of the automatic reconstruction models for the $^{12}$CO $J=2$--1 (green) and $^{13}$CO $J=2$--1 (orange) lines are presented in Figure~7.  The velocity-channel maps produced from the automatic reconstruction are almost exactly the same as the observational map; this is not surprising because the model is constructed by scanning directly the observed channel maps.  The close match between the channel maps of the model and the observations tells us that the model is properly created. We note that the $^{13}$CO data used to create the automatic reconstruction is averaged over a 4.4~km~s$^{-1}$ range, which is slightly larger than the integration range used for the $^{12}$CO data (2.0~km~s$^{-1}$). Therefore, the orange feature ($^{13}$CO) in Figure~7 is slightly more enhanced in comparison with the green feature ($^{12}$CO). 

The $^{12}$CO $J=2$--1 and $^{13}$CO $J=2$--1 lines exhibit different intensity distributions in Figure~7. For example, in the velocity range from 16~km~s$^{-1}$ to 28~km~s$^{-1}$, the emission region of the $^{13}$CO $J=2$--1 line is surrounded by the $^{12}$CO $J=2$--1 emission. These differences are presumably due to the difference of the optical depth of the two lines. If this is the case, we can consider that the two lines trace the structure of the CO gas component of the central region in a mutually complementary manner.

Figure~8 shows the three-dimensional visualization of the CO-gas component of the central region in the 24 different directions. As in Figure~7, both the $^{12}$CO (green) and $^{13}$CO (orange) data are combined to create this diagram. The (0, 0) image is the one as seen along the line of sight. According to Figure~8, the structure of the CO gas component of the central region seems to be a patchy torus, and no high-velocity jets are found. The inner and outer radii and thickness of the torus are  $\sim$1.8, 6.2 and 4.4\arcsec, respectively. When  translated into linear sizes, these angular sizes correspond to  $2.6\times10^{16}$~cm, $9.1\times10^{16}$~cm and $6.5\times10^{16}$~cm at 980~pc. The apparent symmetry axis of the torus is almost the same as that found in the ellipsoidal-shell model (see, Table~1). In the edge-on views of the torus [for example, ($\Theta$, $\Phi$)$=$(45$^{\circ}$, $-$90$^{\circ}$), (45$^{\circ}$, $-$45$^{\circ}$)], we also find non-negligible amount of CO gas located in the polar regions.  The  CO structure may therefore be interpreted as an ellipsoidal shell with multiple holes in the polar region rather than as a patchy torus. This CO structure resembles the H$_2$ structure derived in \citep{cox02}. 

We note that the $z$-direction of the model can be freely squeezed or stretched without altering the intensity distribution [the $z$-direction is the direction perpendicular to the page in the (0, 0) image in Figure~8]. The reason why we have assumed $v=2r''$ here is that the proportionality factor of 2 gives the most symmetric distribution of particles in the three-dimensional space.


\section{Structure of the Central Region and the Existence of High-Velocity Jets}

High-velocity outflows are common in PNe. They are responsible for the creation of the high-density shells seen in PNe, and the interaction between these shells and the fast wind lead to high-temperature shocked gas that lead to thermal X-ray emission.  The detection of X-rays in NGC 7027 \citep{kas94} as well as other PNe such as NGC 6543 \citep{chu01} and BD~+30$^{\circ}$3639 \citep{kas00} has been attributed to this mechanism.  In the latter two examples, their respective fast winds have been detected directly through UV spectroscopy \citep{leu96, per89}.  No fast outflow has been directly detected in NGC 7027, probably because of its high dust extinction.  Shocks have also been suggested to explain the overabundance of the N$_2$H$^{+}$ molecule and the underabundance of CS and HNC in NGC~7027 \citep{zha08}.

The existence of multipolar nebulae have led to the suggestion that the fast outflows in PNe may not be spherically symmetric but are highly collimated in certain directions.  The structure of the molecular shell \citep{gra93a,kas94,cox97} around the ionized gas in NGC 7027 has been suggested as passages of multiple collimated high-velocity jets  \citep{cox02}. If we assume that the jets create ``holes'' in the central region, we may be able to retrace the trail of the unseen jets based on the structural information of the central region. This method was first applied for NGC~7027 by \citet{cox02}, who suggested position-angles of three possible jets, based on the structural information of the H$_2$ emission region. Here, we revisit this discussion by adding the present CO data to the previous H$_2$ and Br$\gamma$ data. Since NGC~7027 is an ionization-bounded nebula \citep[see, e.g.,][]{rod09}, the emission regions of the Br$\gamma$, H$_2$ and CO lines construct a stratified structure. Therefore, by adding the new information obtained by the present CO observation, we can get better structural information of the central region.

In order to compare the three datasets of the Br$\gamma$, H$_2$ and CO lines, we constructed {\it Shape} automatic reconstruction models for the CFHT data of the H$_2$ 1--0 S(1) and Br$\gamma$ lines, in the same manner as the present CO data. The channel maps of the H$_2$ 1--0 S(1) and Br$\gamma$ lines were taken from Figures~3 and 6 in \citet{cox02}. We assumed a constant expanding velocity of 20 km~s$^{-1}$ for the Br$\gamma$ model and a linear velocity-law ($V \propto r$) for the H$_2$ 1--0 S(1) model following the assumptions made by \citet{cox02} in their kinematic modeling. Since the proportional constant for the linear velocity law for the H$_2$ 1--0 S(1) model was not specified in \citet{cox02}, we found an appropriate constant [$v(r)=2.4r''$] through trial and error, so that the model could reproduce the feature seen in Figure~11 in \citet{cox02}. In fact, these velocity laws are not inconsistent with \citet[][at the inner region (roughly within 3--4$''$); since Walsh's law is constructed on the basis of the motion of the central ionized region, it is natural that the difference is large at the outer neutral region]{wal97}.  

In Figure~9, we present the three-dimensional representations of the {\it Shape} model combining the Br$\gamma$, H$_2$ and CO data, and found that the emission regions of the three different lines occupy the spatial positions in a mutually complementary manner, creating a single, patchy ellipsoidal shape (the individual structure of the H$_2$ and Br$\gamma$ emission is given in Figures~12 and 13).

Using the combined {\it Shape} models, we tried to find ``trails'' left behind by the passage of jets. Specifically, we inspected the {\it Shape} particle distributions in various different directions using the {\it Shape} 3D module, and tried to find directions in which the particle distribution has a doughnut-like feature with a hole at the center. In this procedure, we fixed the positions angles at PA=$4^{\circ}$, $-28^{\circ}$ and $-53^{\circ}$, which have been given by \citet{cox02}, and adjusted only the inclination angle (i.e., inclination of the axis to the line of sight). The position angles given by \citet{cox02} are presented in Figure~10, superimposed on the (0, 0) image in Figure~9. We found that the combined model of the three emission lines exhibits a hole only in the direction of (PA, Incl.)=($-53^{\circ}$, $145^{\circ}$) [hereafter, direction A (red bicone in Figures~11, 12 and 13)]. Except for this direction, emission regions of the three emission lines mutually close the angles, and therefore no other holes at other angles were found. 

We then removed the Br$\gamma$ component from the model, and repeated a same inspection to find a doughnut-like feature suggesting the trail of jets. The reason we removed the Br$\gamma$ emission from the model is that the CO and H$_2$ emissions exhibit more patchy structure as compared with the Br$\gamma$ structure (see, Figures~11, 12 and 13), and also that the CO and H$_2$ emission trace the neutral region while the Br$\gamma$ emission traces the ionized region. With the model of just the CO and H$_2$ emission, we found two other viewing angles showing a doughnut-like feature, namely (PA, Incl.)=($4^{\circ}$, $125^{\circ}$) [hereafter, direction B (yellow bicone in Figures~11, 12 and 13)] and (PA, Incl.)=($-28^{\circ}$, $65^{\circ}$) [hereafter, direction C (blue bicone in Figures~11, 12 and 13)]. Particle distributions seen in these directions are presented in Figures~11, 12 and 13. Here, we note that the extended particle distribution of the Br$\gamma$ line (see, Figure~13) is due to the constant velocity law we assumed here. With the constant velocity law, the position of a particle is not uniquely determined in the $z$-direction, and therefore the particle distribution shows the weak, extended tail.

In the top-right panels of Figures~11, 12 and 13, we can clearly see the doughnut-like distributions of the particles, while in the other two directions (see, middle-right and bottom-right panels in Figures~11, 12 and 13) the behavior of the Br$\gamma$ emission is clearly different from that of the CO and H$_2$ emission: the Br$\gamma$ feature shows a hole only in the direction A, while the CO and H$_2$ features show holes in all directions of A, B and C. 

In direction A, all the three components of Br$\gamma$, H$_2$ and CO features show a hole and we can be relatively confident that there has been a passage of a physical jet along that direction. In the directions B and C, however, the Br$\gamma$ emission region has no corresponding holes. 

If we assume a UV-radiation field with a bipolar shape, this might be alternative explanation of the hole-formation, because the UV radiation might create a hole on the structure of CO and H$_2$ components by the effect of photodissociation. However, an important point is that the UV radiation does not seem to create a hole in the structure of Br$\gamma$. In the present observation and modeling, we can clearly see a hole in the Br$\gamma$ structure (see, Figsure~13) in the direction of PA=$-$53$^{\circ}$, and therefore the possibility of the UV radiation must be excluded, at least, in this direction.


\section{Summary}
In this paper we  report the results of an SMA observation of NGC~7027 in the CO $J=2$--1 and $^{13}$CO $J=2$--1 lines. We also performed a morpho-kinematic analysis with {\it Shape}. The main results of this research are summarized below:

\begin{enumerate}
\item The flux of the CO-gas component of the central region is almost fully recovered both in the CO $J=2$--1 and $^{13}$CO $J=2$--1 lines, and the CO structure of the central region is apparent by combining the CO and $^{13}$CO data.
\item The simple ellipsoidal shell model explains the intensity distribution around the systemic velocity, but the high velocity features deviate from this ellipsoidal model.
\item Through the {\it Shape} automatic reconstruction model, we found a possible trail of a jet along the direction of (PA, Incl.)=($-53^{\circ}$, $145^{\circ}$).
\item Holes in other directions are seen in the CO and H$_2$ structure, but these holes are closed by the Br$\gamma$ emission. 
\end{enumerate}

These observations suggest that the central region of NGC 7027 has a torus structure, with the polar directions being cleared away by a high-velocity outflow. As the ionization front propagates into the outer regions, NGC 7027 will develop into a bipolar nebula in emission lines as NGC 6302 and NGC 2346. There is a possibility that there exist high-velocity outflows in other directions, which may lead to the formation of a multipolar nebula. However, in such an interpretation, we should ensure consistency with the present result. Results of this work suggest that molecular-line interferometry observations when interpreted by a morpho-kinematic model can provide valuable insights into the true 3-D structure of PNe.


\acknowledgments
We thank Wolfgang Steffen for discussions on the {\it Shape} program. We also thank Selina Chong, Chih-Hao Hsia and Bosco Yung for their useful comments and stimulating discussions. This research was supported by grants from the Research Grants Council of the Hong Kong Special Administrative Region, China (Project No. HKU 703308P, HKU 702807P and HKU 704209P) and the Seed Funding Programme for Basic Research of the University of Hong Kong (Project No. 200802159006). NK acknowledges the support by the Natural Sciences and Engineering Research Council of Canada, the Killam Trusts and Alberta Ingenuity.



\begin{figure}
\epsscale{.60}
\plotone{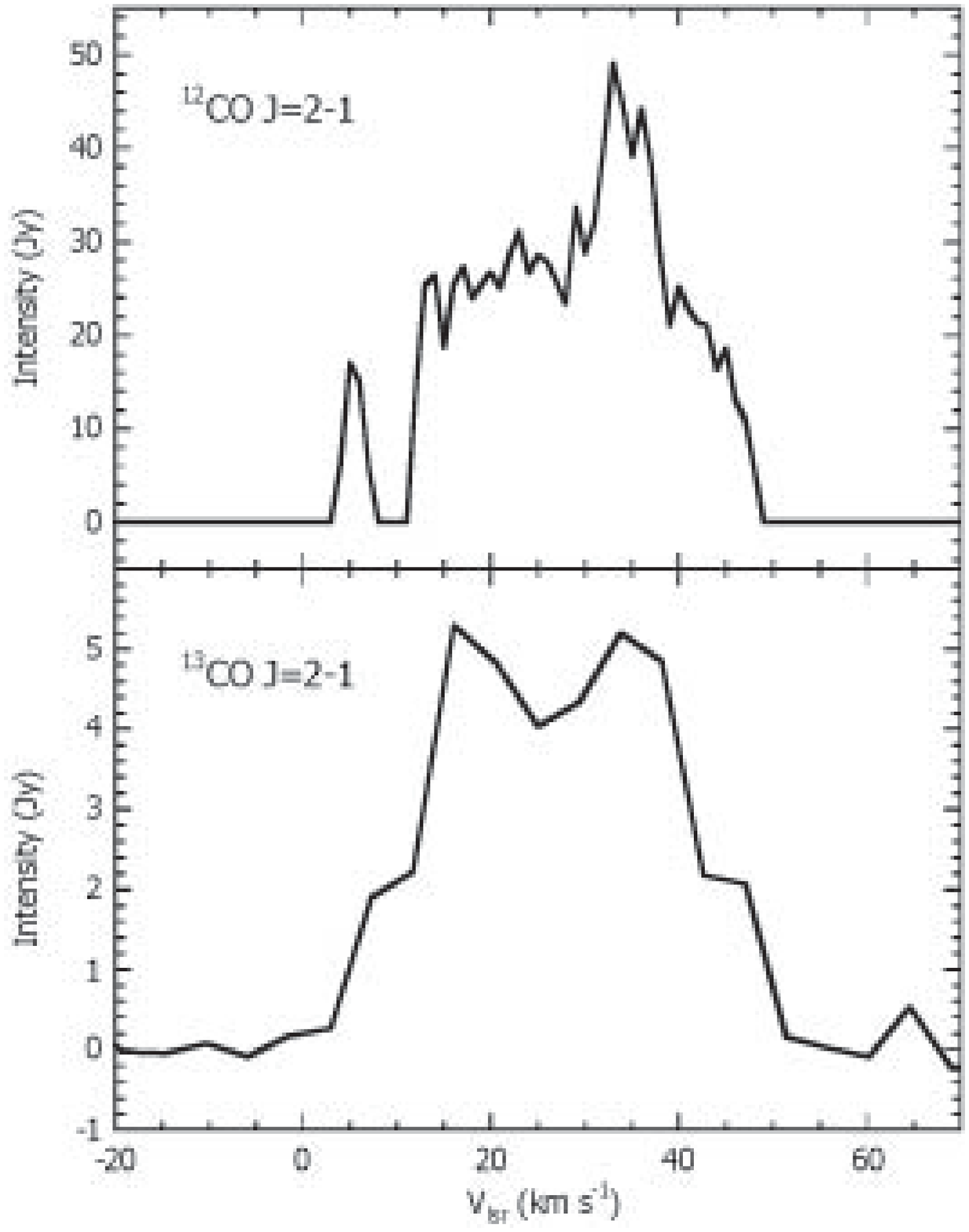}
\figcaption{Total flux line profiles of the $^{12}$CO $J=2$--1 and $^{13}$CO $J=2$--1 lines. \label{fig1}}
\end{figure}
\clearpage

\begin{figure}
\epsscale{.50}
\plotone{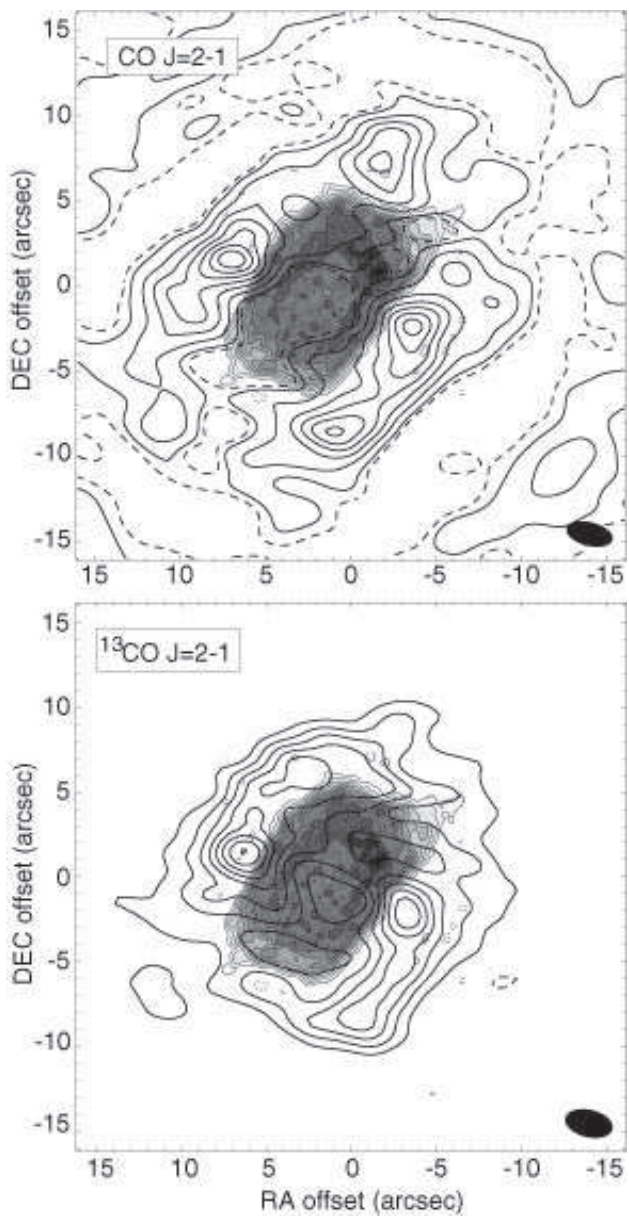}
\figcaption{Total flux intensity maps in the $^{12}$CO $J=2$--1 ({\it upper panel}) and $^{13}$CO $J=2$--1 ({\it lower panel}) lines, superimposed on the $L$-band ($\lambda=$18--20~cm) continuum image \citep[gray scale;][]{bai03}. In the upper panel, contour levels are $-$3.0, 3.0, 15.6, 28.1, 40.7, 53.3, 65.9, 78.4~$\sigma$, and the 1~$\sigma$ level corresponds to $1.5\times10^{-2}$ Jy~beam$^{-1}$. In the lower panel, contour levels are $-$3.0, 3.0, 5.9, 8.9, 11.8, 14.7, 17.6, 20.6, 23.5~$\sigma$, and the 1~$\sigma$ level corresponds to $6.3\times10^{-3}$ Jy~beam$^{-1}$. The FWHM beam size is located in the bottom right corners. The origin of the coordinate corresponds to the phase center. \label{fig2}}
\end{figure}
\clearpage

\begin{figure}
\epsscale{.70}
\plotone{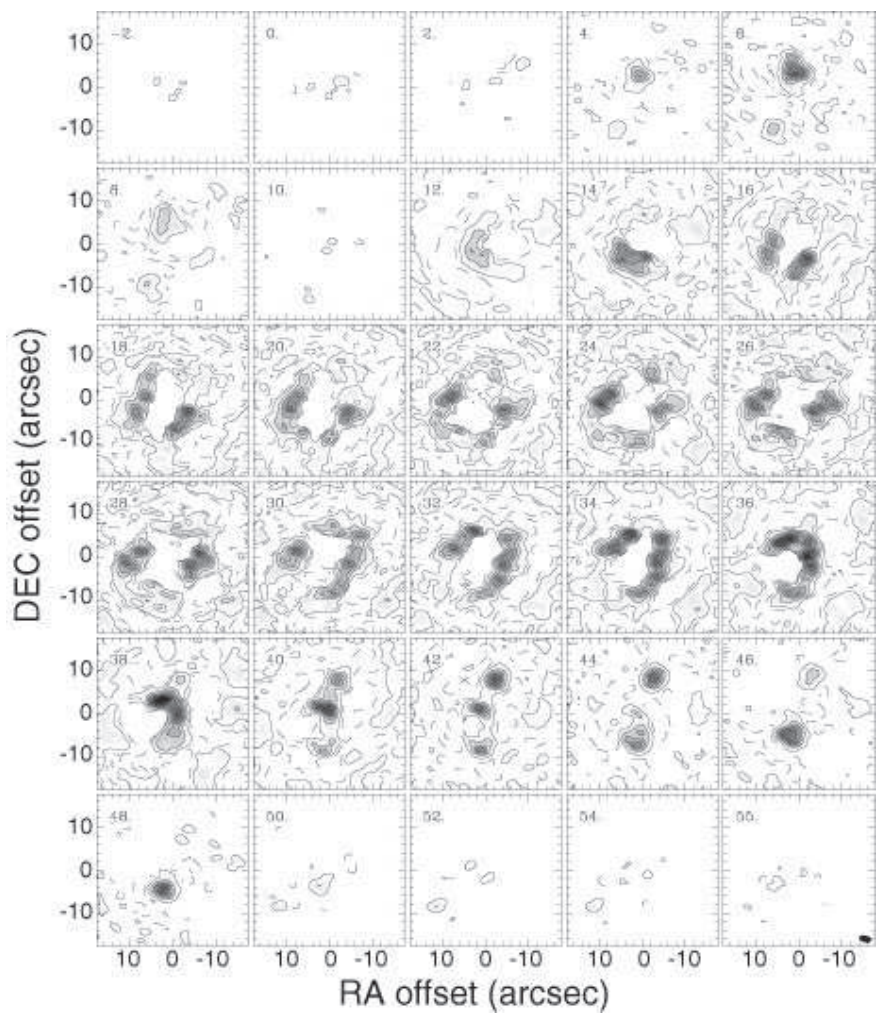}
\figcaption{Velocity channel maps of the SMA compact array data in the $^{12}$CO $J=2$--1 line. The velocity width of each channel is 2 km s$^{-1}$ and the central velocity in km s$^{-1}$ is located in the top left corner of each channel map. The contours levels are $-3$, 3, 14.71, 26.43, 38.14, 49.86, 61.57, 73.29, 85~$\sigma$, and the 1~$\sigma$ level corresponds to $8.04\times10^{-2}$~Jy~beam$^{-1}$. The dashed contour corresponds to $-$3~$\sigma$. The FWHM beam size is located in the bottom right corner of the last channel map. The origin of the coordinate corresponds to the phase center. \label{fig3}}
\end{figure}
\clearpage

\begin{figure}
\epsscale{.70}
\plotone{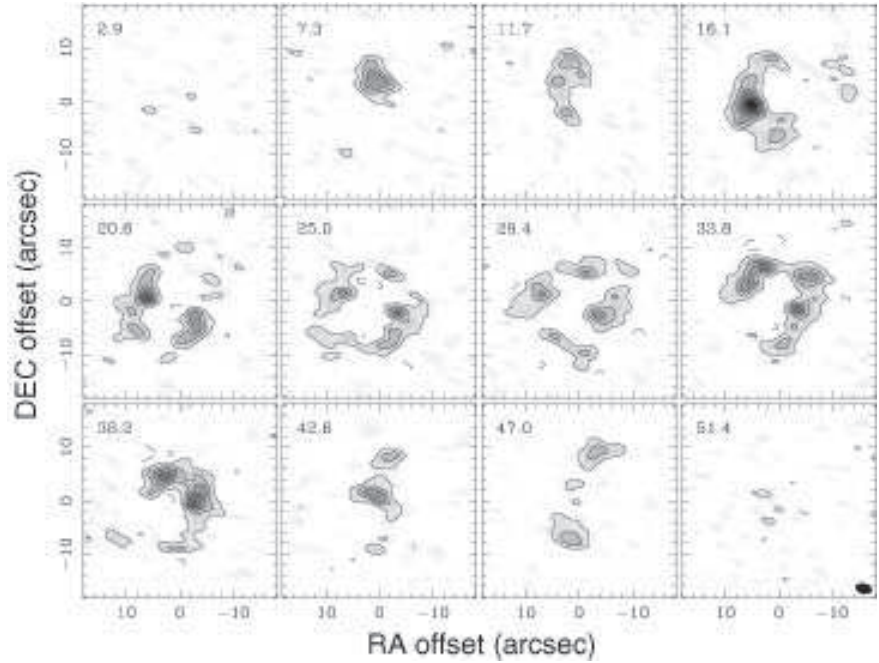}
\figcaption{Velocity channel maps of the SMA compact array data in the $^{13}$CO $J=2$$-$$1$ line. The velocity width of each channel is 4.4~km~s$^{-1}$ and the central velocity in km~s$^{-1}$ is located in the top left corner of each channel map. The contours levels are $-$3.0, 3.0, 7.7, 12.4, 17.1, 21.8, 26.5 and 31.2, and the 1~$\sigma$ level corresponds to $2.19\times10^{-2}$~Jy~beam$^{-1}$. The dashed contour correspond to $-$3~$\sigma$. The FWHM beam size is located in the bottom right corner of the last channel map. The origin of the coordinate corresponds to the phase center. \label{fig4}}
\end{figure}
\clearpage

\begin{figure}
\epsscale{.80}
\plotone{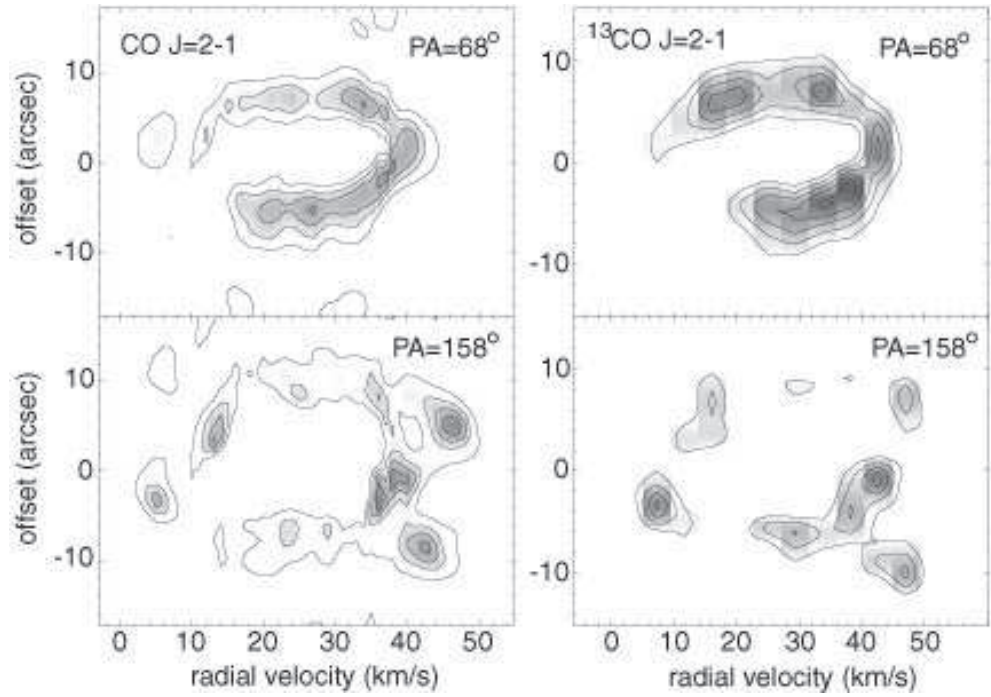}
\figcaption{Position-velocity diagrams of the CO $J=2$--1 (left panels) and $^{13}$CO $J=2$--1 (right panels) lines. The cuts passing the origin have position angles of 68$^{\circ}$ and 158$^{\circ}$. The contours in the left panels are 0.34, 1.49, 2.64, 3.79, and 4.94~Jy~beam$^{-1}$. The contours in the right panels are 0.066, 0.116, 0.166, 0.216 and 0.266~Jy~beam$^{-1}$. \label{fig5}}
\end{figure}
\clearpage

\begin{figure}
\epsscale{.90}
\plotone{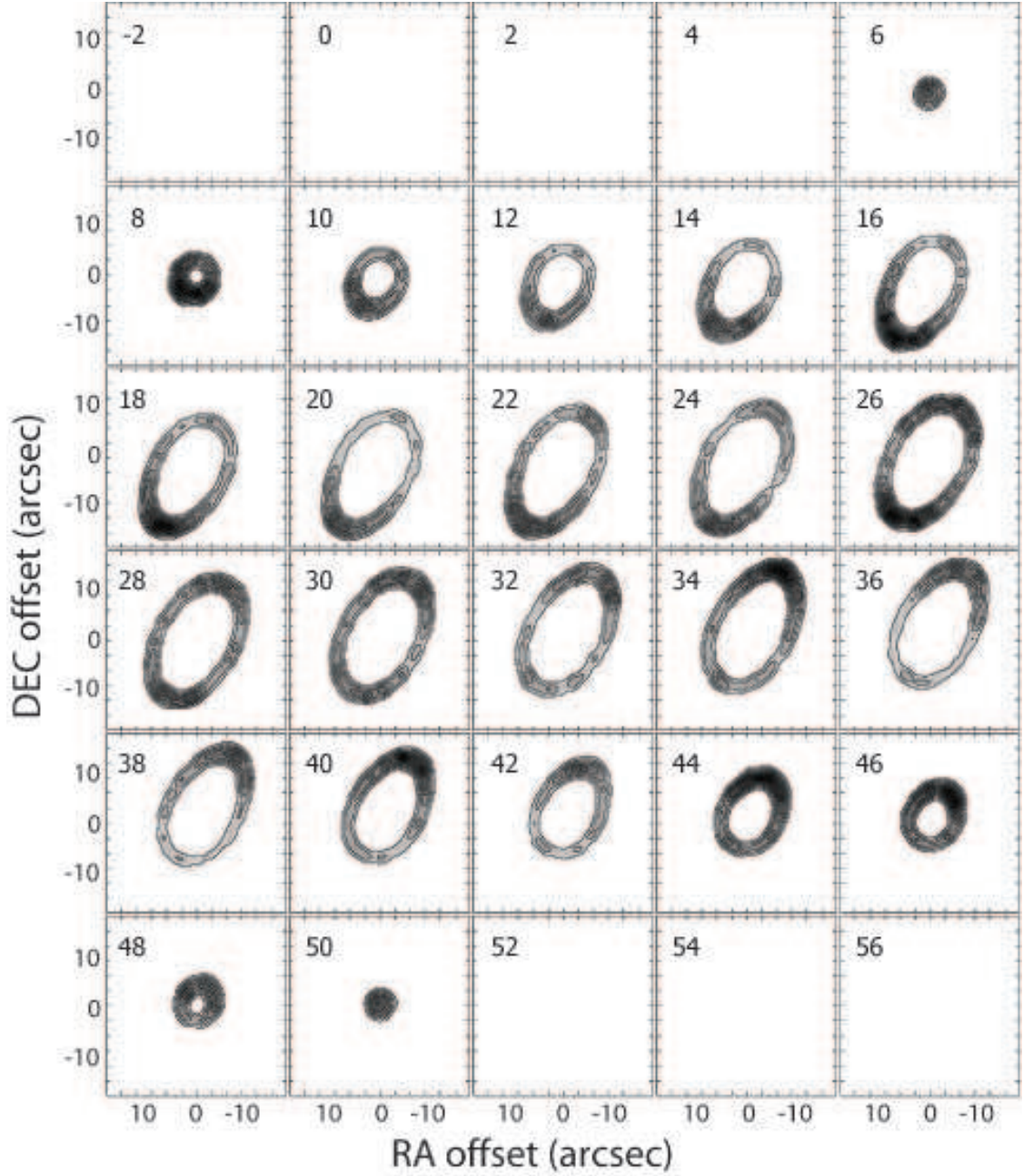}
\figcaption{Channel maps of the best-fit ellipsoidal shell model. For simulating image convolution, a circular beam with a diameter of $2''$ is used. \label{fig6}}
\end{figure}
\clearpage

\begin{figure}
\epsscale{.90}
\plotone{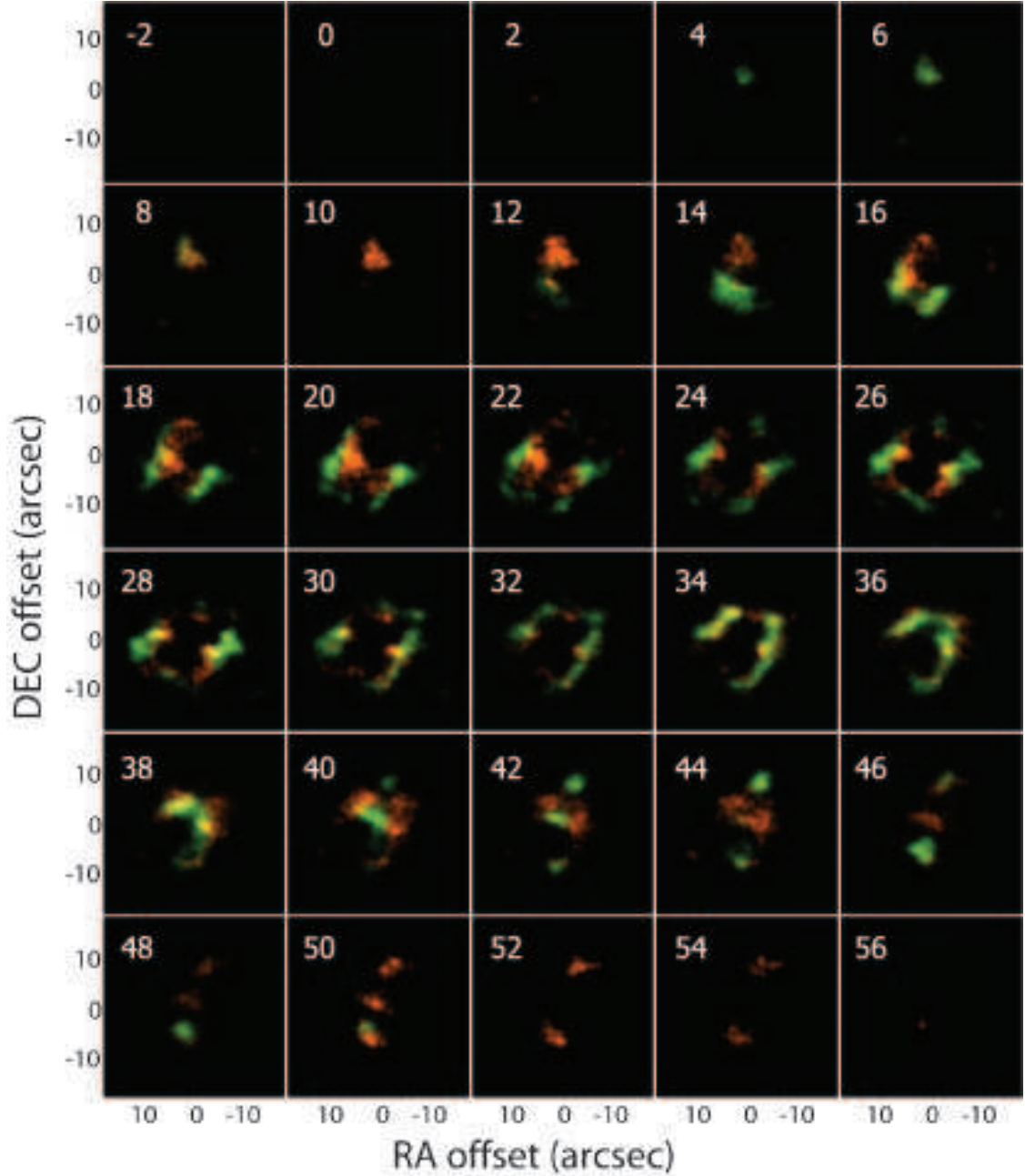}
\figcaption{Channel maps of the automatic reconstruction model. The green and orange colors respectively represent the $^{12}$CO $J=2$--1 and $^{13}$CO $J=2$--1 lines. For simulating image convolution, a circular beam with a diameter of $2''$ is used. \label{fig7}}
\end{figure}
\clearpage

\begin{figure}
\epsscale{.70}
\plotone{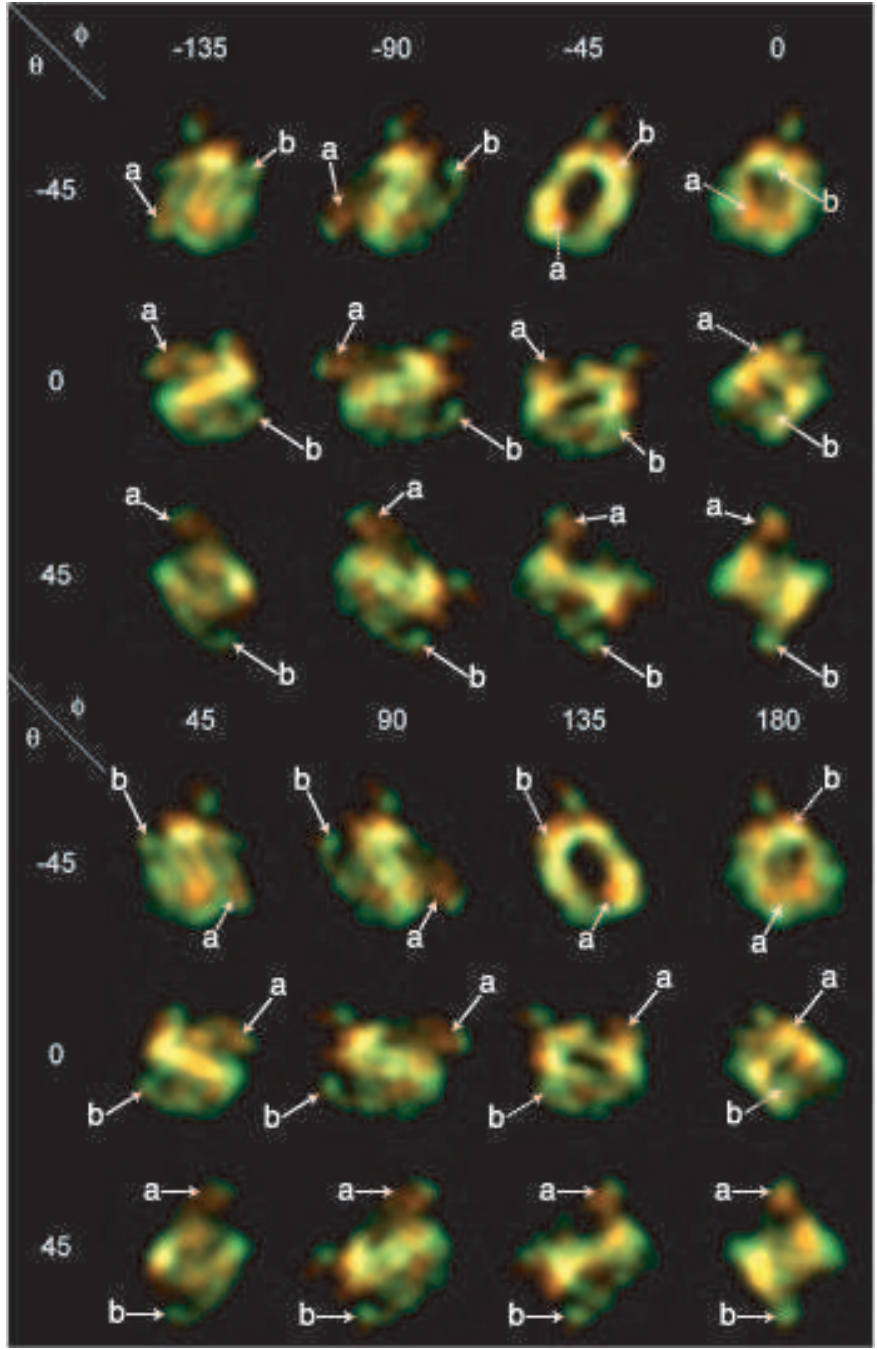}
\figcaption{Three-dimensional representation of the molecular core of NGC~7027 in the $^{12}$CO $J=2$--1 and $^{13}$CO $J=2$--1 lines as seen from 24 directions separated by 45$^{\circ}$. The line of view is identified by ($\Theta$, $\Phi$), where $\Theta$ is the zenith angle and $\Phi$ is the azimuthal angle. The (0, 0) image is the nebula as seen for the Earth. The green and orange colors respectively represent the $^{12}$CO $J=2$--1 and $^{13}$CO $J=2$--1 lines. The arrows indicate representative features. If the feature is lying in the far side and is obscured by another foreground feature, the position of the hidden feature is indicated by the dotted arrows. \label{fig8}}
\end{figure}
\clearpage

\begin{figure}
\epsscale{.70}
\plotone{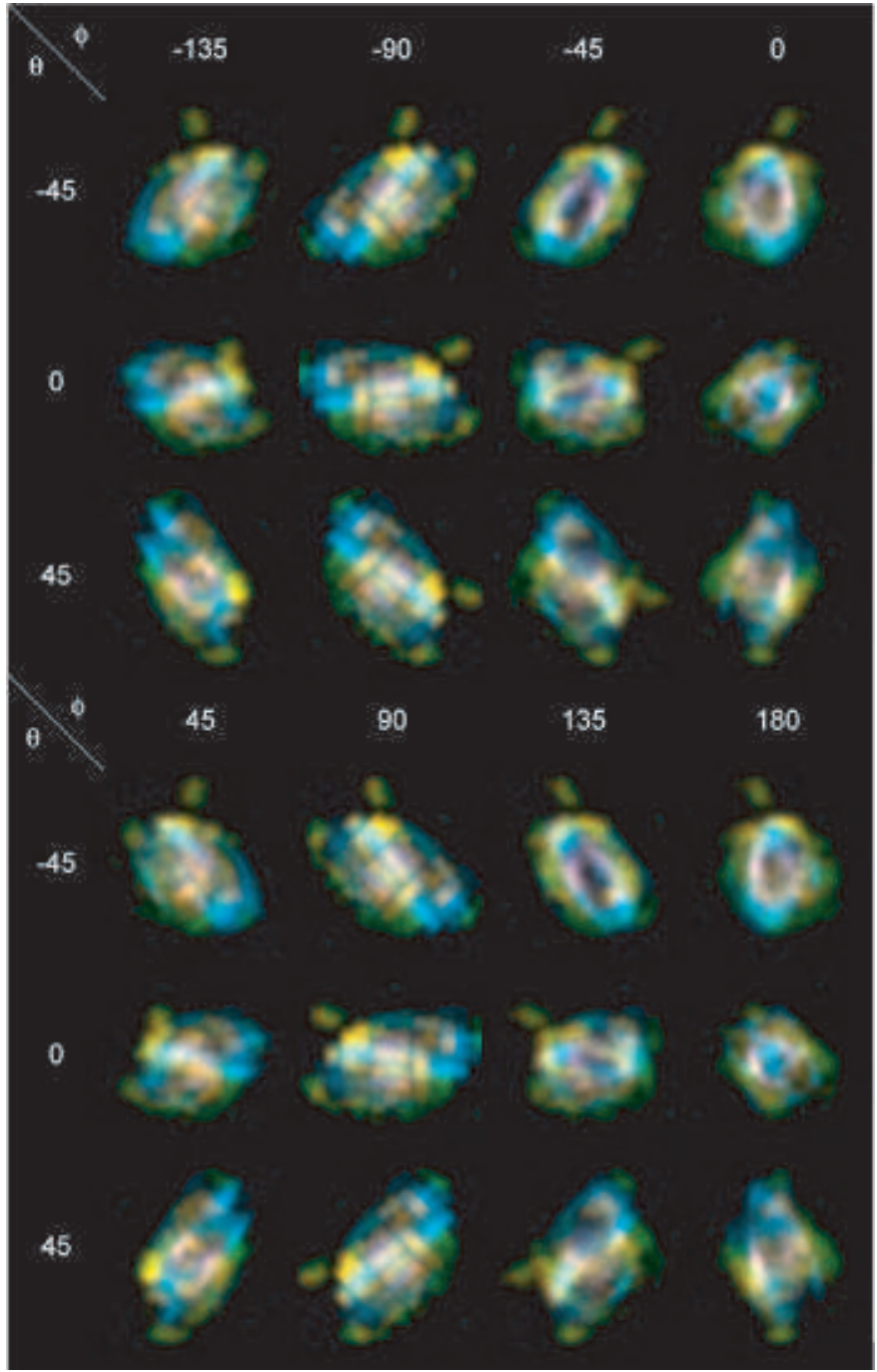}
\figcaption{Three-dimensional representation of the emission regions of the CO [$^{12}$CO $J=2$--1 and $^{13}$CO $J=2$--1], H$_2$ [1--0 S(1)] and Br$\gamma$ lines. The notation of the line of view is as in Figure~8. The CO, H$_2$ and Br$\gamma$ lines are represented by the yellow, blue and white colors, respectively (Note: both the $^{12}$CO and $^{13}$CO are represented by the same yellow color). \label{fig9}}
\end{figure}
\clearpage

\begin{figure}
\epsscale{.60}
\plotone{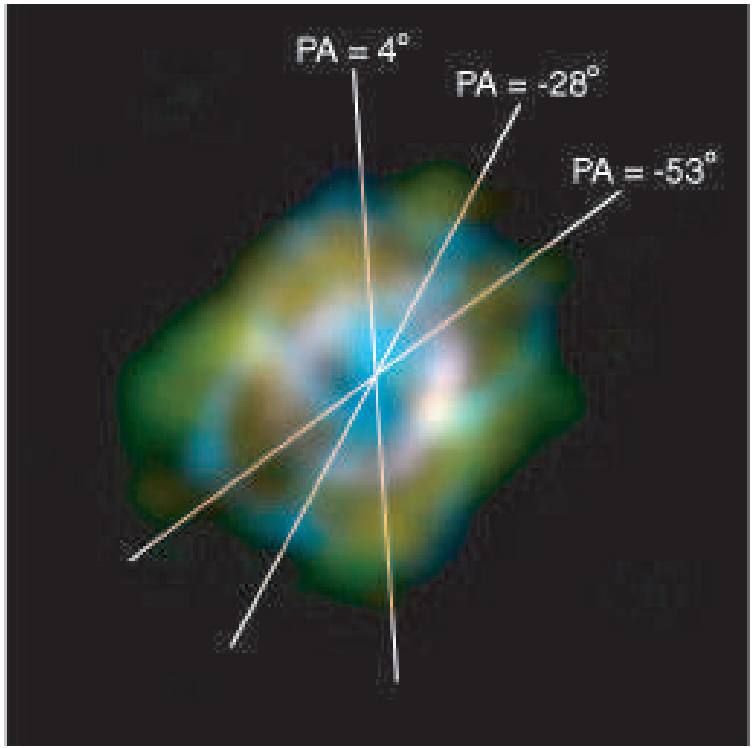}
\figcaption{Position-angles of three possible jets as suggested by \citet{cox02} superimposed on the (0, 0) image in Figure~9. \label{fig10}}
\end{figure}
\clearpage

\begin{figure}
\epsscale{.90}
\plotone{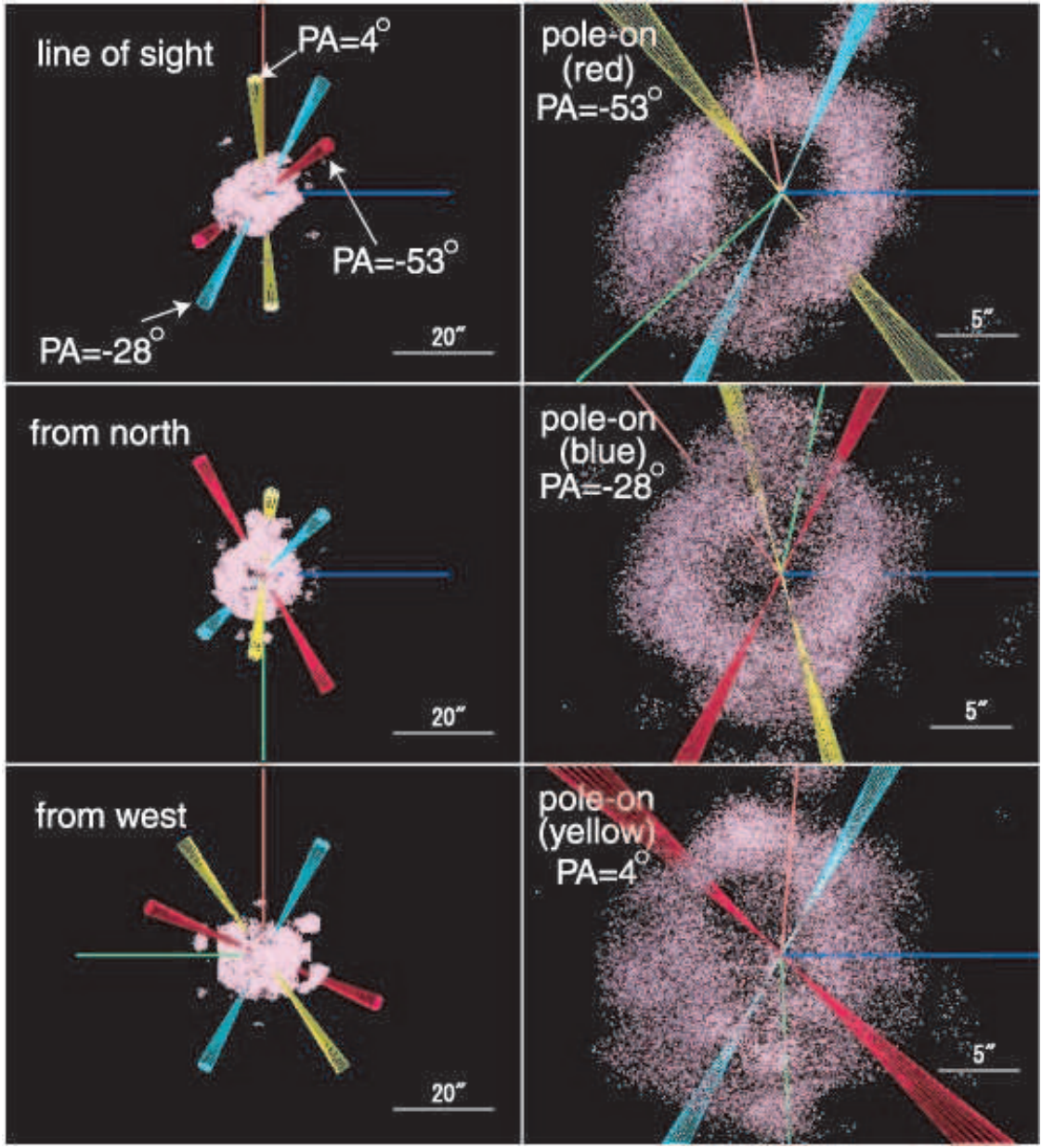}
\figcaption{Three-dimensional location of the three possible jets (yellow, blue and red bicones) seen from six different directions, embedded in the {\it Shape} model of the CO lines (pink particles). The CO data includes both the $^{12}$CO and $^{13}$CO lines. Inclination angles (inclination to the axis to the line-of-sight) of the yellow, blue and red bicones are 125$^{\circ}$, 65$^{\circ}$ and 145$^{\circ}$, respectively. Lengths and opening angles of the bicones are arbitrary. \label{fig11}}
\end{figure}
\clearpage

\begin{figure}
\epsscale{.90}
\plotone{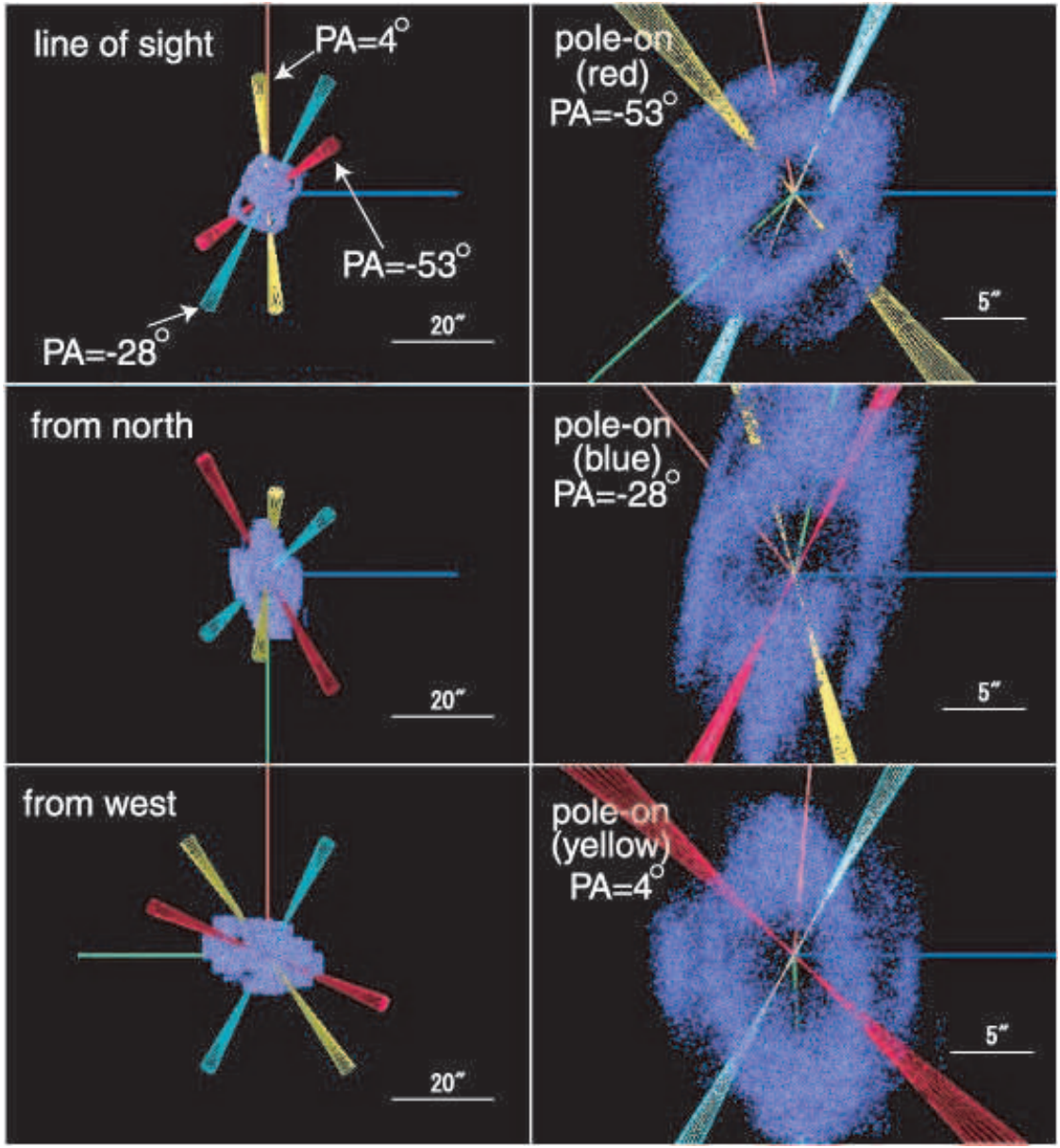}
\figcaption{Similar diagrams to Figure~11 about the H$_2$ 1--0 S(1) line (blue particles). \label{fig12}}
\end{figure}
\clearpage

\begin{figure}
\epsscale{.90}
\plotone{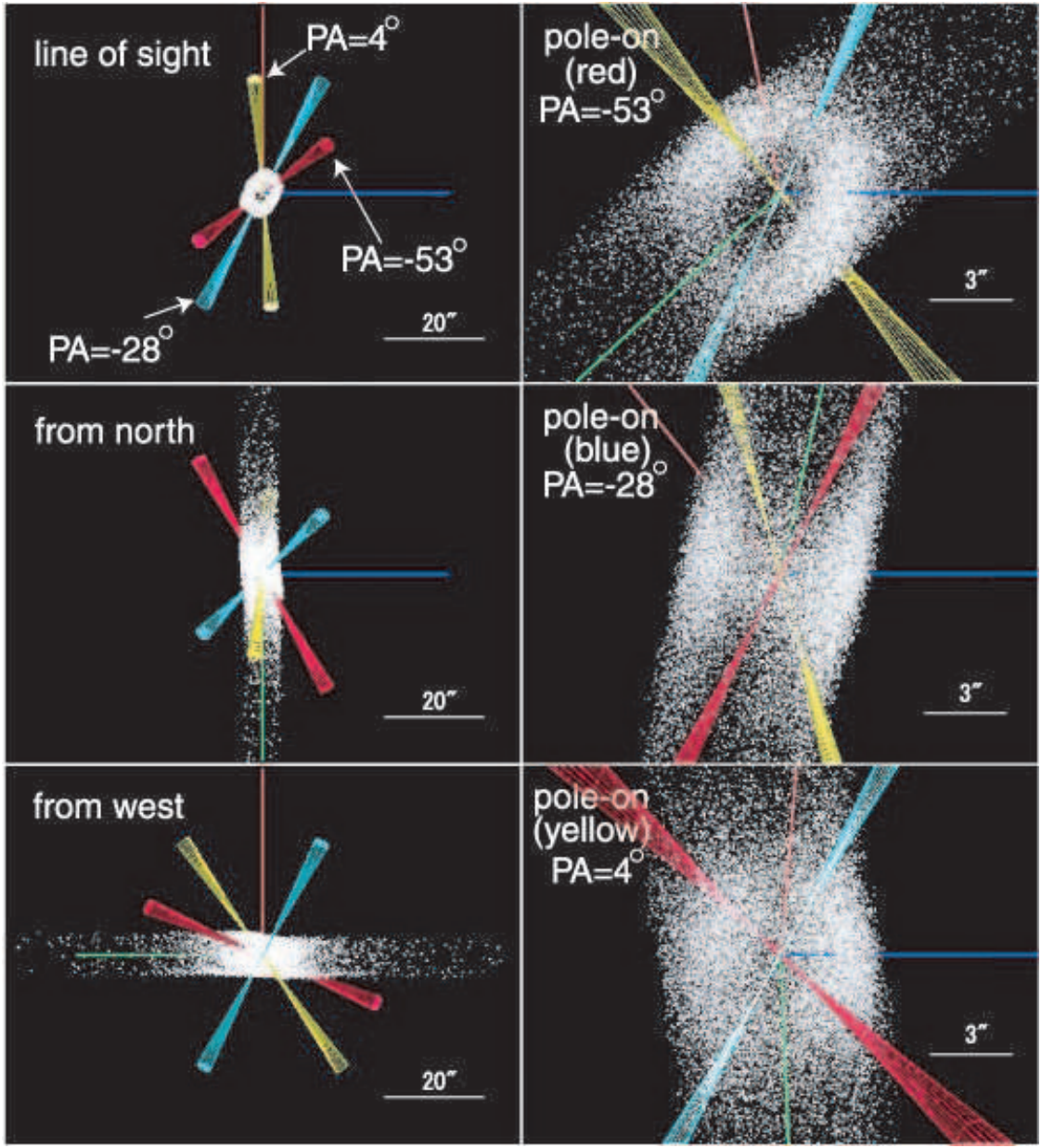}
\figcaption{Similar diagrams to Figure~11 about the Br$\gamma$ line (white particles). The elongated feature is not real (see, text). \label{fig13}}
\end{figure}
\clearpage


\begin{deluxetable}{ll}
\tablecolumns{2}
\tablewidth{0pc}
\tablecaption{Ellipsoidal Shell Model of the Molecular Core}
\tablehead{
\colhead{Parameter} & \colhead{Value} }
\startdata
Inclination of major axis to the line of sight & 60$^{\circ}$ \\
Position angle of major axis & 155$^{\circ}$ \\
Major axis radius & 16$''$ \\
Minor axis radius & 7$''$ \\
Thickness of the shell & 3$''$ \\
System velocity (LSR) & 25.6 km~s$^{-1}$ \\
Expansion velocity & 22 km~s$^{-1}$ \\
\enddata
\end{deluxetable}

\clearpage

\end{document}